\documentclass[11pt,showpacs,amsmath,amssymbl]{revtex4}

\usepackage{amssymb}
\usepackage{graphicx}
\usepackage{epsfig}
\usepackage{multirow}
\usepackage{epstopdf}
\usepackage{gensymb}
\usepackage{nicefrac}
\usepackage{mathtools}
\usepackage{siunitx}
\usepackage{color}

\newcommand{\buck}{C$_{60}$~}
\usepackage[left]{lineno}
\begin{document}

\title{Spatially-resolved, substrate-induced rectification in C$_{60}$ bilayers on copper}
\author{J. A. Smerdon$^1$, P. Darancet$^2$, J. R. Guest$^2$}
\affiliation{$^1$Jeremiah Horrocks Institute of Mathematics, Physics and Astronomy, University of Central Lancashire, Preston, PR1 2HE, UK}

\affiliation{$^2$Center for Nanoscale Materials, Argonne National Laboratory, Argonne IL 60439, USA}

\date{\today}


\begin{abstract}
We demonstrate rectification ratios ($RR$) of $\gtrsim$1000 at biases of 1.3~V in bilayers of \buck deposited on copper.  Using scanning tunneling spectroscopy and first-principles calculations, we show that the strong coupling between \buck and the Cu(111) surface leads to the metallization of the bottom \buck layer, while the molecular orbitals of the top \buck are essentially unaffected.  Due to this substrate-induced symmetry breaking and to a tunneling transport mechanism, the system behaves as a hole-blocking layer, with a spatial dependence of the onset voltage on intra-layer coordination. Together with previous observations of strong electron-blocking character of pentacene/\buck bilayers on Cu(111), this work further demonstrates the potential of strongly-hybridized, \buck-coated electrodes to harness the electrical functionality of molecular components. 
\end{abstract}

\maketitle

The submitted manuscript has been created by UChicago Argonne, LLC, Operator of Argonne National Laboratory (``Argonne"). Argonne, a U.S. Department of Energy Office of Science
laboratory, is operated under Contract No. DE-AC02-06CH11357. The U.S. Government retains for itself, and others acting on its behalf, a paid-up nonexclusive, irrevocable worldwide
license in said article to reproduce, prepare derivative works, distribute copies to the public, and perform publicly and display publicly, by or on behalf of the Government. The Department
of Energy will provide public access to these results of federally sponsored research in accordance with the DOE Public Access Plan. http://energy.gov/downloads/doe-public-access-plan

\newpage

Going beyond the use of molecules as merely resistive elements is central to creating higher electrical functionality in organic optoelectronics devices. 
The earliest proposal for such non-linear electrical behavior is the Aviram-Ratner molecular diode model, proposed in 1974~\cite{Aviram:1974eh}. Following theoretical models of large rectification~\cite{taylor2002theory,Andrews08-jacs}, much experimental work has been done in optimizing the intrinsic molecular properties, level alignment, and electronic coupling to electrodes~\cite{dhirani97-jcp,diezperez09-nc,kushmerick02-prl,selzer02-jpcb,guedon2012observation,batra2013tuning,Kim14-pnas,capozzi2015single,randel2014unconventional,perrin2016gate}. While rectification ratios in excess of 100 have been reported for ionically-\cite{capozzi2015single} or electrostatically-gated~\cite{perrin2016gate} molecules, the electrical performance of such molecular devices remains several orders of magnitude below that of their inorganic counterparts.   A primary reason is that, despite their high tunability, molecules are very sensitive to their immediate environment, so that much of their desirable intrinsic electrical properties are lost when integrated into actual devices due to the spectral broadening of the molecular levels. Minimizing such effects leads to an apparent paradox, as it implies the physical decoupling of the electrodes from the active region of the device, which dramatically degrades its electrical performance.

In a previous work~\cite{smerdon2016large}, we showed this paradox can be entirely overcome by using metallized molecules~\cite{LuCrommie} as a buffer layer between the active device region and the metallic electrode. Specifically, the strong coupling between \buck and the (111) surface of Cu leads to good electron and hole injection from the metal (due to a strongly broadened density of states around the Femi energy), while limiting interactions between the metal and molecules deposited in a second layer. In~\cite{smerdon2016large}, using a donor pentacene monolayer on top of \buck/Cu(111), we synthesized electron-blocking layers with average rectification ratios in excess of 1000 at 1 V, where the current flow is suppressed for a positively-biased sample.
 
In this work, we demonstrate large rectification ratios in bilayers of \buck deposited on copper, where the symmetry between the layers is broken by the strong  \buck/~Cu(111) interaction. Using scanning tunneling microscopy and spectroscopy (STM/STS), we show that dense, close-packed second layers of \buck forms on top of  \buck/~Cu(111) and exhibits large rectification ratios. Using density functional theory (DFT), we show that --while the bottom layer of \buck is metallized-- the second layer conserves its original accepting and semiconducting character, leading to rectification. Interestingly, in contrast to pentacene~/~\buck/~Cu(111), the system is a hole-blocking layer ({\it i.e.} current flow is suppressed for a negatively-biased sample), showing the versatility and the potential of the \buck/~Cu(111) system to support non-linear current-voltage characteristics.  
Additionally, spatially-resolved $I(V)$ curves reveal a shift in the `on' voltage of the rectifier between the edge and the `bulk' of the bilayer; this phenomenon is reproduced in the calculations and explained as a Stark effect dependent on the local intra-layer coordination.

\section{Methods}
 Measurements were carried out in a ultrahigh-vacuum (UHV) variable-temperature (VT) STM operating with the sample maintained at 60~K.  The Cu(111) crystal was cleaned by sputtering with Ar$^+$ ions at 1~keV and simultaneously annealing at 900~K, with a final sputtering cycle at room temperature followed by a brief anneal at 900~K. \buck was deposited in the same UHV system using an organic molecular-beam evaporator at  710~K.  This \buck film was annealed to ~$\approx$570~K, either during deposition or immediately after. This process results in the desorption of any \buck molecules not in immediate contact with the Cu substrate.  It also causes Cu atoms to diffuse away from locations beneath \buck molecules such that there is a 7-atom vacancy beneath each molecule~\cite{Pai:2004bc}.  These Cu atoms relocate to nearby step edges, where they tend to attach in such a way as to straighten step edges along close-packed rows, resulting in facetted crystallographic steps. The 2-d structure of the film is $p(4\times4)$. The second \buck layer was deposited in the same way but with the sample at room temperature.

Tungsten tips were prepared by electrochemical etching; when they occasionally adsorbed molecules during experiments, they were cleaned by {\it in situ} UHV heating to approximately 1000~K to restore imaging and spectroscopic quality.  Bias voltage $V_B$ was applied to the sample.  The $I(V)$  curves were obtained with a `grid' spectroscopy approach, where the measurements were obtained sequentially in a grid pattern over the relevant region of the sample.  Due to our high sensitivity and low noise in the regions of `zero' measured current, capacitive coupling contributed measurable offsets (due to the voltage ramp during spectroscopy) and oscillating backgrounds (due to $V_B$ modulation for $dI/dV$ measurements), which were removed during data analysis.

The first-principles DFT calculations were performed using the generalized gradient approximation (GGA) of Perdew, Burke, and Ernzerhof (PBE)~\cite{PBEGGA}, as implemented in the SIESTA package~\cite{siesta}. We used standard Troullier-Martin pseudopotentials for the description of C and Cu (respectively 4 and 11 electrons) taken from the Siesta pseudopotential database. The electronic density was represented on a $300$~Ry grid and converged  to $10^{-5}$ number of electrons. Brillouin zone integrations were performed using an electronic temperature of $0.05$~eV a $4\times4\times1$ ($4\times2\times1$) $k$-point grid for the full (half) coverage calculations. Our structural approximant consists of two \buck molecules deposited on a 6-layer-thick slab of Cu (16 Cu atoms per layer, hcp binding site) in an hexagonal unit cell (with a 10.21~\AA~in-plane lattice parameter, and a 60~\AA~lattice parameter in the direction perpendicular to the surface, corresponding to $\simeq$27~\AA~of vacuum between the top-most C atom and the bottom layer of Cu). Interatomic forces were optimized below 0.04~eV/\AA, while keeping the 4 bottom-most Cu layers in their bulk positions. Electronic density and density of states were computed using a slab dipole correction along the $z$ axis.

\section{Results and discussion}

\begin{figure}
\includegraphics[width=1\textwidth]{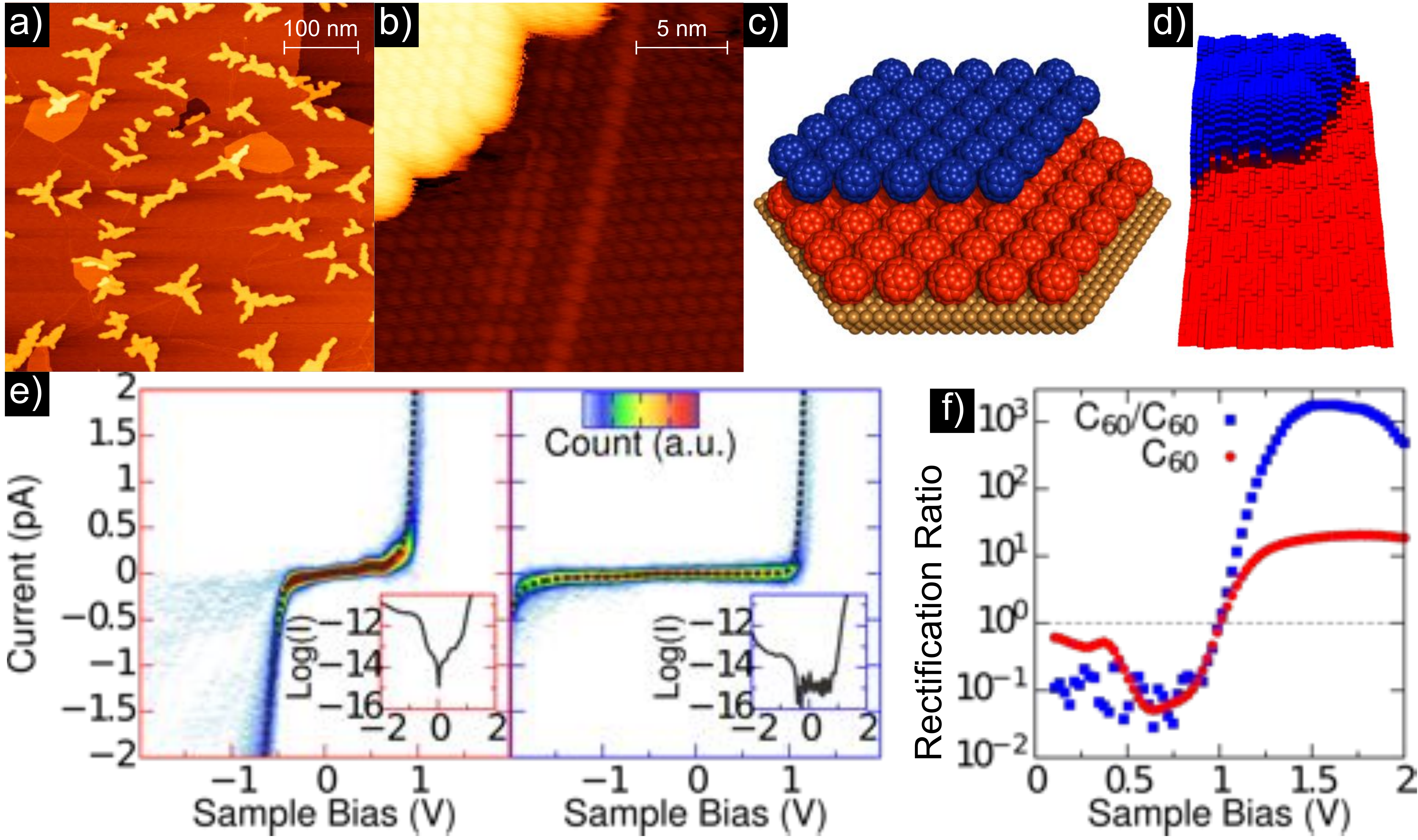}
\caption{(a) Scanning tunneling topograph of the \buck/Cu(111) bilayer (\SI{500}{\nano\meter}, $V_B=1.6$~V, $I_T=0.3$~nA). (b) Detail of the film (\SI{20}{\nano\meter}, $V_B=1.63$~V, $I_T=0.22$~nA). (c) A model of the film showing an ideal arrangement.  The red molecules are seated on an hcp site of Cu(111), the blue molecules are in hollow sites atop the red molecules. (d) Grid spectroscopy topograph.  The data are shown as square columns, each representing a $I(V)$ curve. (e) Histograms of the $I(V)$ curves from the red (monolayer) area (left) and the blue (bilayer) area (right). (f) Rectification ratios ($RR = \left| \nicefrac{I(V)}{I(-V)} \right| $) of the average current-voltage characteristics for the monolayer (blue) and bilayer (red) areas.}
\label{1}
\end{figure}

In Figure~\ref{1}, we show the arrangement of the \buck film grown as described above. The large-scale STM topographic image in Fig.~\ref{1}(a) shows what appears to be dendritic growth, where close-packed directions are slightly preferred for growth, leading to large fern-shaped islands.  Additionally, most islands appear to be located adjacent to or on top of a translational domain boundary, which are imaged as extended linear features (again, oriented along close-packed directions). Figure~\ref{1}(b) shows a detail of the topography, with an extended translational domain boundary in the monolayer running up the middle of the image and a second-layer \buck island visible in the top left corner.  Molecules at the edge of the second-layer \buck film appear larger than molecules within the island due to convolution with the shape of the tip.  In Fig.~\ref{1}(c) we present a model of a second layer \buck island atop a monolayer of \buck atop fully-reconstructed Cu(111).

We explored the transport characteristics and spatial dependence of this molecular system by measuring 2112 $I(V)$ curves over a 10$\times$5.2~nm$^2$ region corresponding to the edge of the bilayer, as indicated in Fig.~\ref{1}(d), where the topography is represented by square columns for each $I(V)$ curve.  By `simultaneously' performing the measurements on both the \buck  monolayer and the bilayer, we were able to confirm that the same tip condition was maintained for both regions; no data was filtered from this set. The tip height at each point was established at the tunneling conditions $V_B=+2.0$~V and $I_{t}=200$~pA.  The data from the monolayer (red) and the bilayer (blue) are separated by height and compiled into histograms on the left and right of Fig.~\ref{1}(e), respectively; averages of the curves are overlaid as a dotted line and shown on a log scale in the inset.  The rectification ratio (RR) ($RR = \left| \nicefrac{I(V)}{I(-V)} \right| $) is computed from the average $I(V)$s and shown in Fig.~\ref{1}(f) and serves as a figure of merit for the degree of rectification of the two molecular systems.

The monolayer of \buck (left, Fig.~\ref{1}(e)) on Cu shows quasi-ohmic behavior for $\left| V_B \right| \leq 0.5$~V, with nonlinear increases in current arising at $\simeq-0.5$~V and $\simeq0.7$~V as new conductance channels appear.  The associated RR dips below 0.1 at lower voltages, but increases to unity at $V_B \simeq$1~V and settles at $\simeq$20 at higher voltages.  In contrast, the $I(V)$ characteristics in the \buck / \buck / Cu(111) region are strongly non-ohmic (right, Fig.~\ref{1}(e)).  As can be seen in the log plot in the inset, for $-0.5$~$\leq V_B \leq 0.9$~V, the measured current is below the instrument noise.  A weak current channel appears at $V_B \simeq$-0.5~V, but the current remains below 100~fA until $V_B \simeq$-1.7~V; as this voltage aligns with the strong conductance channel in the \buck monolayer, this weak current may be due to direct tunneling to that \buck monolayer.  At positive bias, a strong current channel turns on at $V_B \simeq$1~V.  As a result of this asymmetry, the RR for this bilayer crosses unity at $V_B \simeq$1~V and achieves $\simeq$2000 for 1.4~$\leq V_B \leq$~1.75~V.  This $100\times$ improvement in the RR is directly a result of the current suppression at negative biases.  Note that the preferred path of electrons -- from acceptor to metallized molecular layer -- is exactly the opposite of our previous observation, which used a donor molecule (pentacene) in the second layer~\cite{smerdon2016large}.

\begin{figure}
\includegraphics[width=0.5\textwidth]{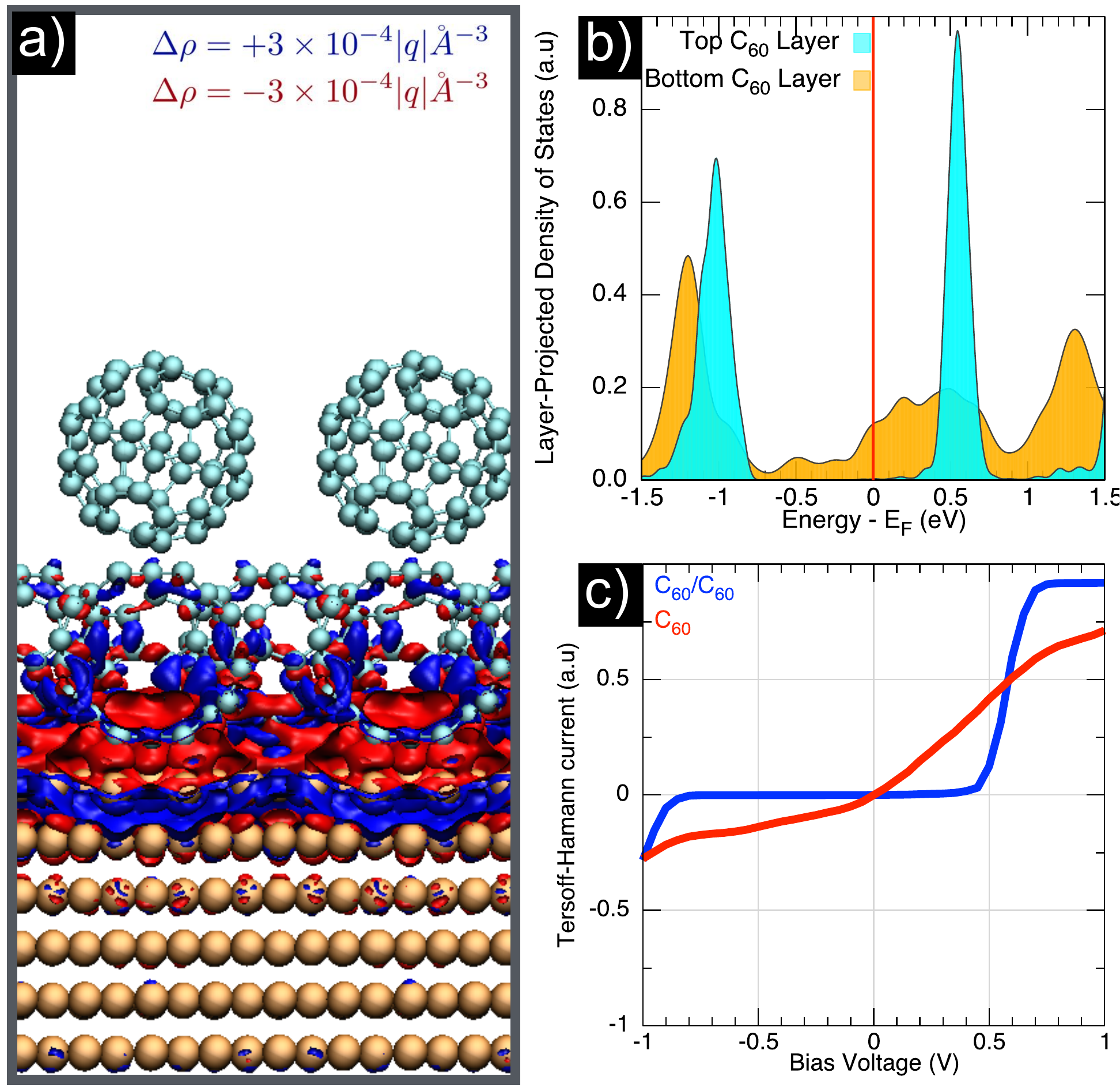}
\caption{Electronic structure of \buck /~\buck /~Cu(111). (a) Isosurface of electronic density difference ($\Delta\rho$) between \buck/~\buck/~Cu(111) and its individual components ($\Delta\rho_{\textrm{C}_{60}/\textrm{C}_{60}/\textrm{Cu(111)}} - \Delta\rho_{\textrm{C}_{60}/\textrm{C}_{60}} - \Delta\rho_{\textrm{Cu(111)}} $). The change in electronic density is localized to the first \buck layer and the two topmost Cu layers. (b) Kohn-Sham density of states (DoS) of \buck/~\buck/~Cu(111) projected on the two \buck layers. Upon binding, the bottom \buck layer has a metallic DoS, while the top \buck layer remains semiconducting.  (c) Corresponding $I(V)$ characteristics using Tersoff-Hamann theory and the Kohn-Sham DoS (the Cu-DoS is assumed to be independent of energy in the energy range considered). The computed $I(V)$ characteristic of the bilayer is diodic, due to the top \buck layer preserved n-type semiconducting character and the continuous DoS in the bottom layer.}
\label{2}
\end{figure}

%%% theory
To understand the origin of the rectifying behavior, we turn to first-principles calculations. As pointed out in previous works~\cite{wang2004rotation,Pai:2004bc, Pai:2010dx},  \buck monolayers on Cu(111) are strongly hybridized, leading to a metallic density of states (DoS) and large increase in their local polarizability. 
As seen in Fig.~\ref{2}, in the case of the \buck bilayer, we find that this behavior applies solely to the layer in contact with Cu(111) and does not impact the topmost layer. First, the change in electronic density upon binding (Fig.~\ref{2} (a)) is entirely localized to the bottom layer. Moreover, as shown in Fig.~\ref{2}(b), this leads to a system with a DoS reminiscent of Schottky-diode, in which the top layer remains a semiconductor with an electron-accepting character, while the bottom \buck layer is a metal. Interestingly, despite the original energy alignment between the lowest unoccupied molecular orbitals (LUMO) of the two \buck layers, the Kohn-Sham LUMO of the top layer is found to sit at 0.58eV above the Fermi energy. We attribute this energy mismatch to the interface dipole at the \buck /~Cu interface, which increases the electrostatic potential~\cite{natan,selloni} above the negatively-charge \buck layer. We note that this effect has also been observed for smaller charge transfer in the case of \buck bilayers on Au(111)~\cite{GrobisCrommie,geng2012}. The increase of the electrostatic potential also suggests that molecules on top of \buck /~Cu will have their highest occupied molecular orbital (HOMO) pushed towards the Fermi energy, in agreement with our experimental findings that the HOMO of pentacene lies 0.25 eV below the Fermi energy on top of \buck/~Cu~\cite{smerdon2016large} -- which may suggest very low onset voltages are achievable for electron-blocking layers on top of \buck/~Cu. 

The functional form of the $I(V)$ characteristics can be understood as resulting directly from the unusual DoS of the system using the same hypotheses as in the pentacene~/~\buck/~Cu system \cite{smerdon2016large}. Specifically, in Fig.~\ref{2} (c) we plot the Laudauer current obtained when the transmission function is the product of the layer-projected DoS ($I\left(V\right)\propto \int^{V}_{0} dE \textrm{DoS}_{\textrm{top}} \left(E\right)\textrm{DoS}_{\textrm{bottom}} \left(E\right)$, where V is the potential of the tip and E the energy). This formula neglects bias-induced changes in the DoS (see below) and possible direct coupling between (1) the top \buck layer and Cu and (2) the tip and the bottom \buck layer. The calculated current has the same diodic lineshape as the one observed in experiments with a smaller onset bias (0.58 V as compared with $\simeq$1.0 V found experimentally) that we attribute to the erroneous Kohn-Sham level alignment (because the onset voltage in the forward direction is related to the energy difference between the frontier molecular orbital and the Fermi energy of the system). Finally, we note that the spectrally-broad LUMO of the bottom \buck layer effectively acts in a manner similar to a \textit{gateway} state~\cite{widawsky} from the point of view of the top \buck layer. While these states were found to dominate the low-bias $I(V)$ for conjugated and non-conjugated molecules~\cite{widawsky}, the weak coupling between the two \buck layers results in a different behavior.     

\begin{figure}
\includegraphics[width=1\textwidth]{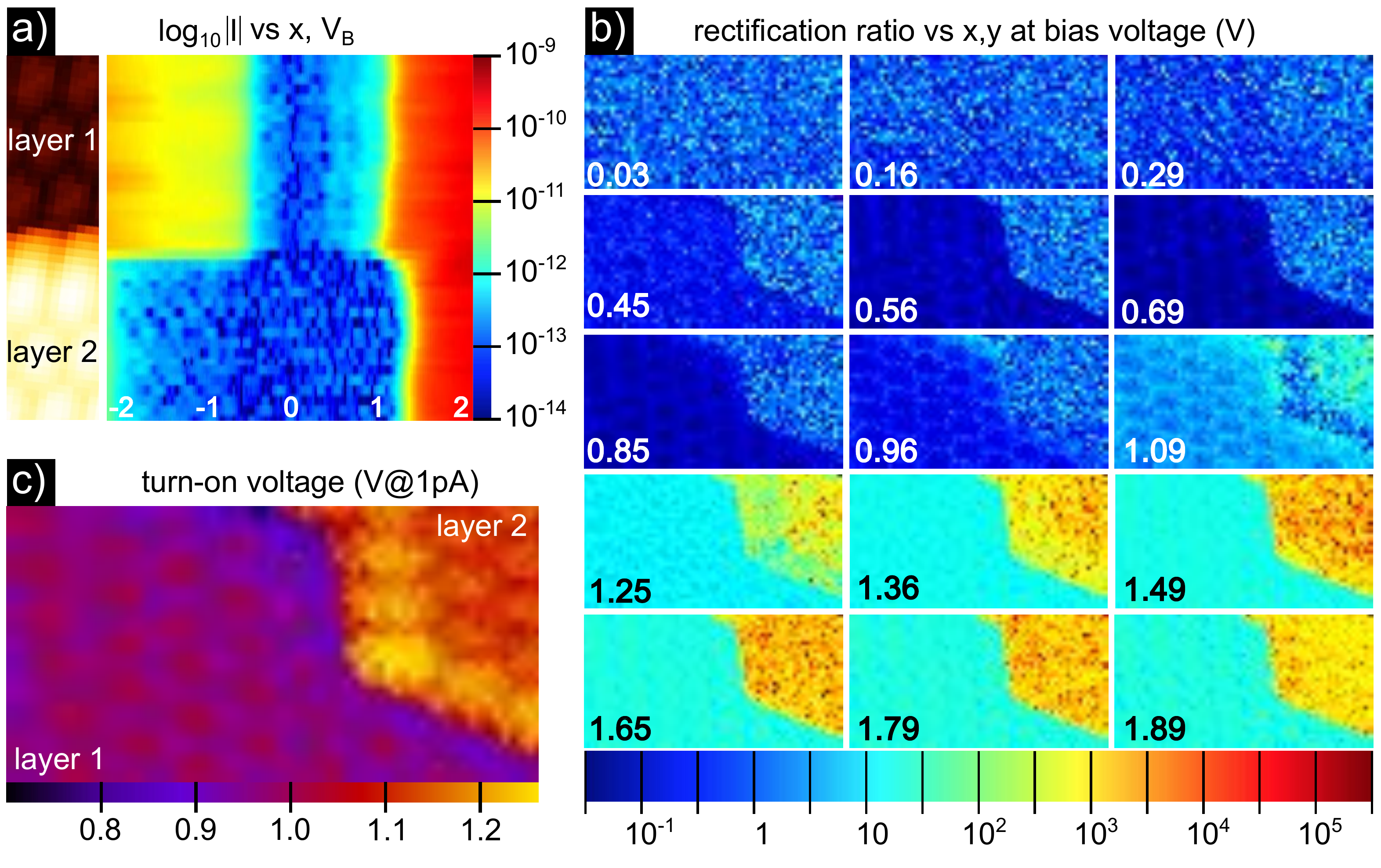}
\caption{Scanning tunneling spectroscopy of \buck/~\buck/~Cu(111). (a) log$_{10}$$|I(V)|$ traveling across a step between the bilayer and the monolayer, with lateral averaging over 10 points. (b) $RR$ maps for the entire bias range, showing zero rectification at low bias and strong rectification for biases above 1.2 V.  The strongest rectification is at 1.6 V. (c) Map of `turn-on' voltage for the system, showing that the turn-on voltage decreases with increasing coordination of the \buck molecule. }
\label{3}
\end{figure}

We now explore the spatial and coordination dependence of our observations. In Fig. \ref{3}, we show data derived from the grid spectroscopy presented in Fig. \ref{1}. The variation in the local $I(V)$ at the interface of the bilayer and monolayer is shown in Fig. \ref{3} (a), where the data is spatially-averaged over 10 points parallel to the edge of the bilayer. In contrast with the metallic behavior of the \buck monolayer, and in agreement with our theoretical findings, the bilayer shows a larger 'turn-on' voltage as well as reduced current in the reverse direction. The local variation of the rectification ratio is shown in Fig. \ref{3} (b). Consistent with the average rectification analysis, the local rectification shows that the monolayer is weakly rectifying with a maximum $RR$ of about 20, while the bilayer is strongly rectifying with local maximum $RR$ up to about $10^4$. Interestingly, we find that, while the rectification ratio is homogeneous in the bilayer, the turn-on voltage is strongly dependent on the distance from the edge of the island and the local intra-layer coordination ({\it e.g.} $V_B$=1.09~V, Fig. \ref{3} (b))~\cite{smerdon2016large}. 

Such behavior is particularly noticeable around 1~V, where the edge of the bilayer starts rectifying. A summary of this phenomena is shown in Fig. \ref{3} (c), which shows the spatial dependence of the turn-on voltage (as defined by the voltage needed for the current to be larger than 1 pA). Such turn-on voltage varies from ~1 V  for the \buck molecules within the island (2d coordination 6) to 1.2 V and 1.25 V for those on the edges (2d coordination 4) and the molecule at the corner (2d coordination 3).  

\begin{figure}
\includegraphics[width=0.5\textwidth]{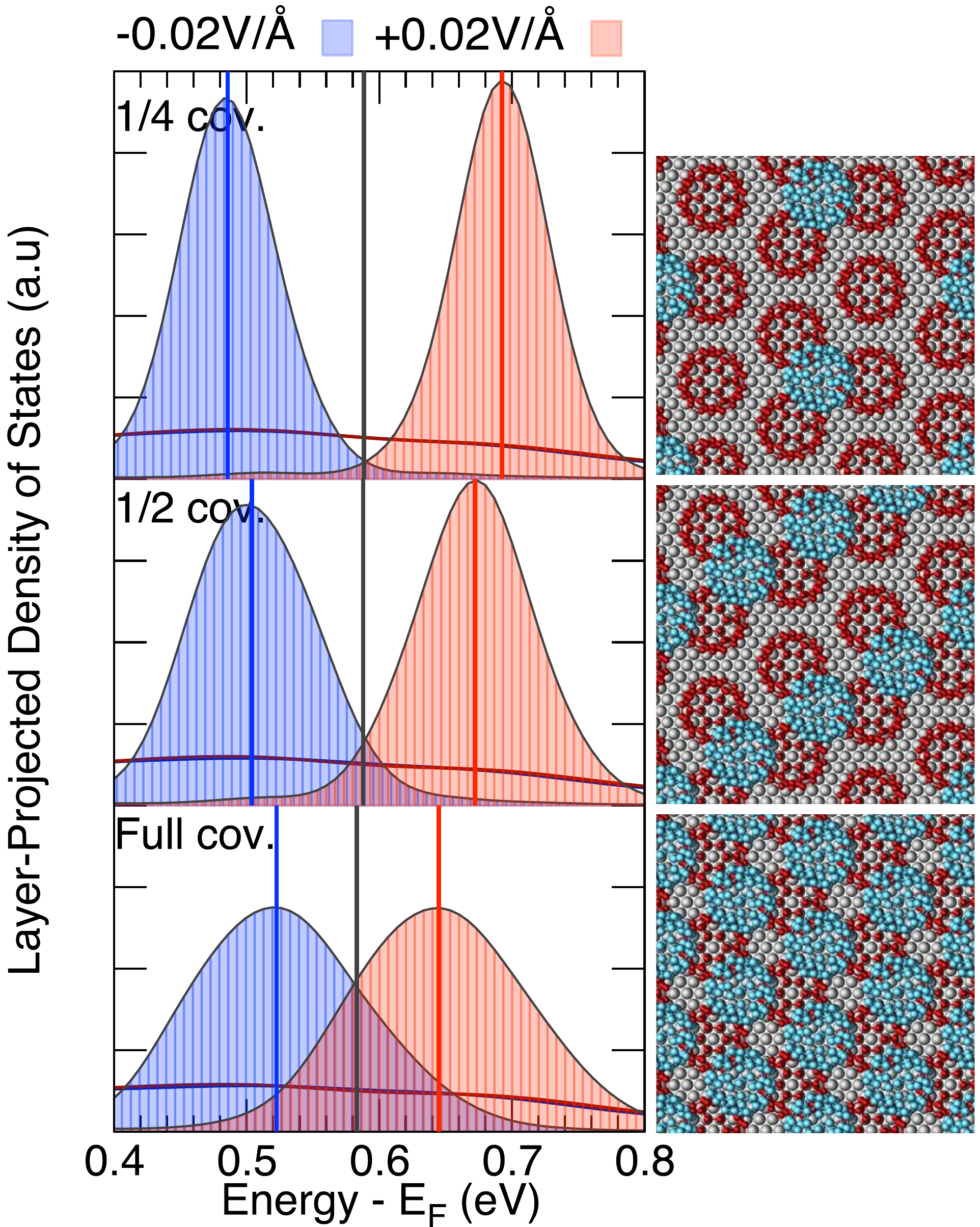}
\caption{Coverage-dependent Stark effect in \buck/~\buck/~Cu(111). Layer-projected DoS from DFT of molecules at quarter (top panel), half (middle), and full (bottom) coverage for the top \buck layer. Red and blue  curves correspond to the DoS under fields of opposite polarities for the top  (filled curves) and bottom (plain lines) layers. Only the top layer DoS is affected by the electric field. Blue, black, and red vertical lines correspond to the maxima of the Gaussian fit of the top layer DoS for the negative, 0-, and positive field cases, respectively. Larger orbital Stark shifts are found at lower coverage. Corresponding structural models used in the DFT calculations are shown on the right.}
\label{4}
\end{figure}

While the number of atoms involved prevents a direct DFT calculation of the electronic structure of the edge and corner molecules, we examine the differences between the fully covered bilayer of \buck on Cu(111) and the half- and quarter-coverage situations (coverage referring to the second \buck layer). In figure~\ref{4}, no significant change is found in the top layer's LUMO energies for the three coverages (within 0.01 eV of each other, as determined by a Gaussian fit of the layer-projected density of states). However, we observe more significant differences in the molecular orbital energies response to an external electric field, \textit{i.e.} a larger orbital Stark effect as coordination is reduced. Specifically, when applying electric fields of $\pm$0.02 V/\AA~ perpendicular to the surface (Fig. \ref{4}), we find the energy shift of the LUMO to be {70\% and 38\% larger in the quarter- and half-coverage cases ($\pm$103 meV and $\pm$83 meV, respectively} as compared with $\pm$61 meV for the full coverage case) due to the change in local dielectric environment. While the structural and field models both differ from the experimental {conditions}, the computed magnitude of this effect is compatible with the experimental observation of a 15-20\% increase of the `turn-on' voltage at the edges. {Moreover, as pointed out in reference~\cite{FernandezTorrente} in the context of \buck monolayers and molecules on Au(111), changes in local dielectric environment are also conducive to impacting the level alignment through non-local static correlations -- an effect beyond the reach of standard DFT approaches~\cite{Neaton}. Both Stark effect and level alignment renormalization result in the same qualitative behavior: the measured LUMO energy increases at lower coordination. Further STS studies on the dependence of the measured LUMO energies on the tip-substrate distance should allow to determine the relative magnitude of these two mechanisms.}

In conclusion, we have observed large rectification ratios in non-covalent, self-assembled, molecular bilayers of \buck on Cu(111). Using DFT, we have shown that, due to its strong coupling with Cu(111), the bottom \buck layer acts as a metallic buffer layer, preserving the semiconducting character of the top \buck layer while allowing electron and hole injection. We have analyzed the local variation of the rectification and found that \buck molecules with smaller coordinations have larger turn-on voltages, a phenomenon we attribute to an enhanced electronic Stark effect in a local environment with reduced dielectric screening.

\section{Acknowledgments}

Use of the Center for Nanoscale Materials, an Office of Science user facility, was supported by the U. S. Department of Energy, Office of Science, Office of Basic Energy Sciences, under Contract No. DE-AC02-06CH11357.  Primary support for this work was provided by the Department of Energy Office of Basic Energy Sciences (SISGR Grant DE-FG02-09ER16109).  {We thank our anonymous referees for bringing some references to our attention, and for suggesting that static non-local correlations may explain the coordination-dependent turn-on voltage~\cite{FernandezTorrente}.}

\end{document}